# Heterogeneous back-end-of-line integration of thin-film lithium niobate on active silicon photonics for single-chip optical transceivers

Lingfeng Wu,[1†] Zhonghao Zhou,[2†] Weilong Ma,[1†] Haohua Wang,[1†] Ziliang Ruan,[3] Changjian Guo,[1,5] Shiqing Gao,[1] Zhishan Huang,[1] Lu Qi,[1,4] Jie Liu,[4] Jing Feng,[2] Dapeng Liu,[2*] Kaixuan Chen,[1,5*] and Liu Liu[3*]

[1]Guangdong Provincial Key Laboratory of Optical Information Materials and Technology, South China Academy of Advanced Optoelectronics, Sci. Bldg. No. 5, South China Normal University, Higher-Education Mega-Center, Guangzhou 510006, China

[2]Chongqing United Microelectronics Center Co., Ltd, No. 20 Xiyuan South Street, Shapingba District, Chongqing 401332, China

[3]State Key Laboratory of Extreme Photonics and Instrumentation, College of Optical Science and Engineering, International Research Center for Advanced Photonics, East Bldg. No. 5, Zijingang Campus, Zhejiang University, Hangzhou 310058, China

[4]State Key Laboratory of Optoelectronic Materials and Technologies, School of Electronics and Information Technology, Sun Yat-Sen University, Guangzhou, 510006 China

[5]National Center for International Research on Green Optoelectronics, South China Normal University, Guangzhou 510006, China

[†]These authors contributed equally.

*dapeng.liu@cumec.cn; chenkaixuan@m.scnu.edu.cn; *liuliuopt@zju.edu.cn*

**Abstract:** The explosive growth of artificial intelligence, cloud computing, and large-scale machine learning is driving an urgent demand for short-reach optical interconnects featuring large bandwidth, low power consumption, high integration density, and low cost preferably adopting complementary metal-oxide-semiconductor (CMOS) processes. Heterogeneous integration of silicon photonics and thin-film lithium niobate (TFLN) combines the advantages of both platforms, and enables co-integration of high-performance modulators, photodetectors, and passive photonic components, offering an ideal route to meet these requirements. However, process incompatibilities have constrained the direct integration of TFLN with only passive silicon photonics. Here, we demonstrate the first heterogeneous back-end-of-line integration of TFLN with a full-functional and active silicon photonics platform via trench-based die-to-wafer bonding. This technology introduces TFLN after completing the full CMOS compatible processes for silicon phonics. Si/SiN passive components including low-loss fiber interfaces, 56-GHz Ge photodetectors, 100-GHz TFLN modulators, and multilayer metallization are integrated on a single silicon chip with an efficient inter-layer and inter-material optical coupling. The integrated on-chip optical links exhibit >60 GHz electrical-to-electrical bandwidth and supports 128-GBaud OOK and 100-GBaud PAM4 transmission below forward error correction thresholds, establishing a scalable platform for energy-efficient, high-capacity photonic systems.

**Introduction**

High-performance photonic integrated circuits (PICs), such as optical transceivers, with ultrahigh bandwidth, low energy consumption, and compatibility with complementary metal-oxide-semiconductor (CMOS) foundry processes are highly desirable[1,2]. Silicon photonics has emerged as a leading candidate, offering high integration density, scalable manufacturing, and cost effectiveness. Commercial platforms can already provide process design kits (PDKs) with co-integration of pure Si modulators of >40 GHz bandwidth and Ge photodetectors of >70 GHz bandwidth[3], enabling demonstrations of 256 Gbps/lane parallel[4] or 80×10 Gb/s wavelength division multiplexed links[5]. Although the bandwidth of a Ge photodetector has been pushed to >250 GHz ultimately without compromising other performances[6], Si modulators on the other hand remain a major bottleneck for the scalability of integrated transceivers on a single Si chip. Imposed by carrier dynamics, they suffer from a weak index modulation, resulting in either a high drive voltage or a limited bandwidth. Mach–Zehnder interferometer (MZI) based Si modulators have shown bandwidths over 110 GHz, yet with a half-wave voltage over 54 V[7,8]. Using a ring resonator structure, the drive voltage of a Si modulator can be decreased. However, the bandwidth and working wavelength range would be compromised[9,10].

To overcome this bottleneck, heterogeneous integration has been pursued to introduce materials more suitable for electro-optic (EO) modulation, such as InP[11,12], special polymers[13,14], two-dimensional materials[15,16], plasmonic structures[17,18], barium titanate[19], lead zirconate titanate[20], and thin-film lithium niobate (TFLN)[21-32]. Among these techniques, silicon–TFLN heterogeneous integration has been proven as the most promising one, as TFLN offers a single crystalline, chemical stable, and high refractive index thin-film with an intrinsic Pockels effect. Modulators with >110 GHz bandwidth, sub-1 V half-wave voltage, <1 dB insertion loss, and excellent linearity have been realized on TFLN[33-36]. Recent demonstrations of bonded TFLN on passive Si or SiN circuits have been achieved using die-to-wafer bonding[21-25], wafer-to-wafer bonding[26–29], or micro-transfer printing[30–32], presenting an efficient interlayer coupling and a record modulation performance on a Si chip. However, extending this approach to active Si photonics platforms presents a fundamental challenge. Besides modulators, such an active Si photonics chip should necessarily include Ge photodetectors, multilayer passive structures, multilayer metallization, and normally a thick $SiO_2$ isolation layer above the photonic waveguides, which prevent the sub-micron proximity requirement for an efficient optical coupling between Si and TFLN structures. Consequently, the ultimate goal of integrating TFLN with multi-functioning active silicon photonics has not been achieved yet.

Here, we demonstrate the first heterogeneous integration of TFLN with active silicon photonics. In contrast to conventional heterogeneous integration strategies, where the introduction of TFLN during the front-end process immediately after the Si or SiN photonic waveguide fabrication would inevitably impose constraints on the subsequent silicon photonics processing concerning material compatibility and thermal budget, we introduce a back-end-of-line (BEOL) integration platform via trench-based die-to-wafer bonding of TFLN. The proposed platform retains a complete, unmodified, and CMOS-compatible silicon photonics process with standard PDKs. The bonding and processing of TFLN photonic structures, as well as the top-most metal electrodes, can be fabricated in a separate fab with only a moderate critical dimension (CD) requirement. This integration strategy truly combines the advantages of silicon photonics and TFLN, enabling co-integration of high-performance modulators and photodetectors on a single Si chip. The resulting optical transceiver includes parallel on-chip optical links of 3-dB electrical-to-electrical (EE) bandwidths over 60GHz, and supports data transmissions at 128-GBaud on-off keying (OOK) and 100-GBaud 4-level pulse amplitude modulation (PAM4). This work establishes a CMOS-compatible and high-performance platform for single-chip optical transceivers for data-center interconnects and microwave photonics[37], as well as dense on-chip data links towards optical network-on-chips (NoCs) in future wafer-scale computing[38,39].

## Results
### BEOL integration of TFLN on active Si photonics.

The proposed BEOL integration process relies on heterogeneous die-to-wafer bonding of TFLN dies onto a fully fabricated silicon photonics wafer. To demonstrate the scalability of this integration platform, the silicon photonics wafer incorporates all essential building blocks for optical transceiver PICs, including Si and SiN passive components, fiber couplers, thermo-optic (TO) phase shifters, Ge photodetectors, and multilayer metallization. To enable efficient optical coupling between Si and TFLN while preserving the active Si device stack, trenches are opened in the thick over-cladding oxide layer above the modulation regions prior to bonding, thereby optically exposing the Si waveguides. Diced x-cut LN-on-insulator (LNOI) dies of appropriate size are then flipped and bonded into these trenches using a thin benzocyclobutene (BCB) adhesive layer. This BCB layer compensates for the surface roughness of the etched trench bottoms and thereby ensures a high bonding yield of the LNOI dies. The phase modulation sections are implemented using TFLN ridge waveguides integrated with standard coplanar ground-signal-ground (GSG) push-pull traveling-wave electrodes. Design details are discussed in Supplementary Section I. Vertical adiabatic couplers (VACs) within the trenches enable nearly lossless mode transitions between the Si and TFLN waveguides. These VACs rely solely on inverse tapers in the Si waveguides, without any fine patterning of the TFLN layer. Figure 1a shows a sketch of the heterogeneous integration process, as well as images of the fabricated transceiver chip. The modulation sections, as well as the VACs, are located within the trenches, whereas other Si photonic elements remain outside. Optical input and output are accessed via Si grating couplers or SiN edge couplers. Figures 1b and 1c show a three-dimensional (3D) representative schematic of one on-chip transceiver channel and a cross-sectional view of the platform, respectively. Images of some key parts of the fabricated structure are also illustrated in Figs. 1a and 1d, confirming the effectiveness of the proposed heterogeneous platform. The entire fabrication flow is concisely summarized in Fig. 2, with further details provided in the "Methods" section. Overall, this process establishes a heterogeneous Si photonics platform integrating high-performance TFLN modulators, broadband Ge photodetectors, and low-loss passive components on a single Si chip.

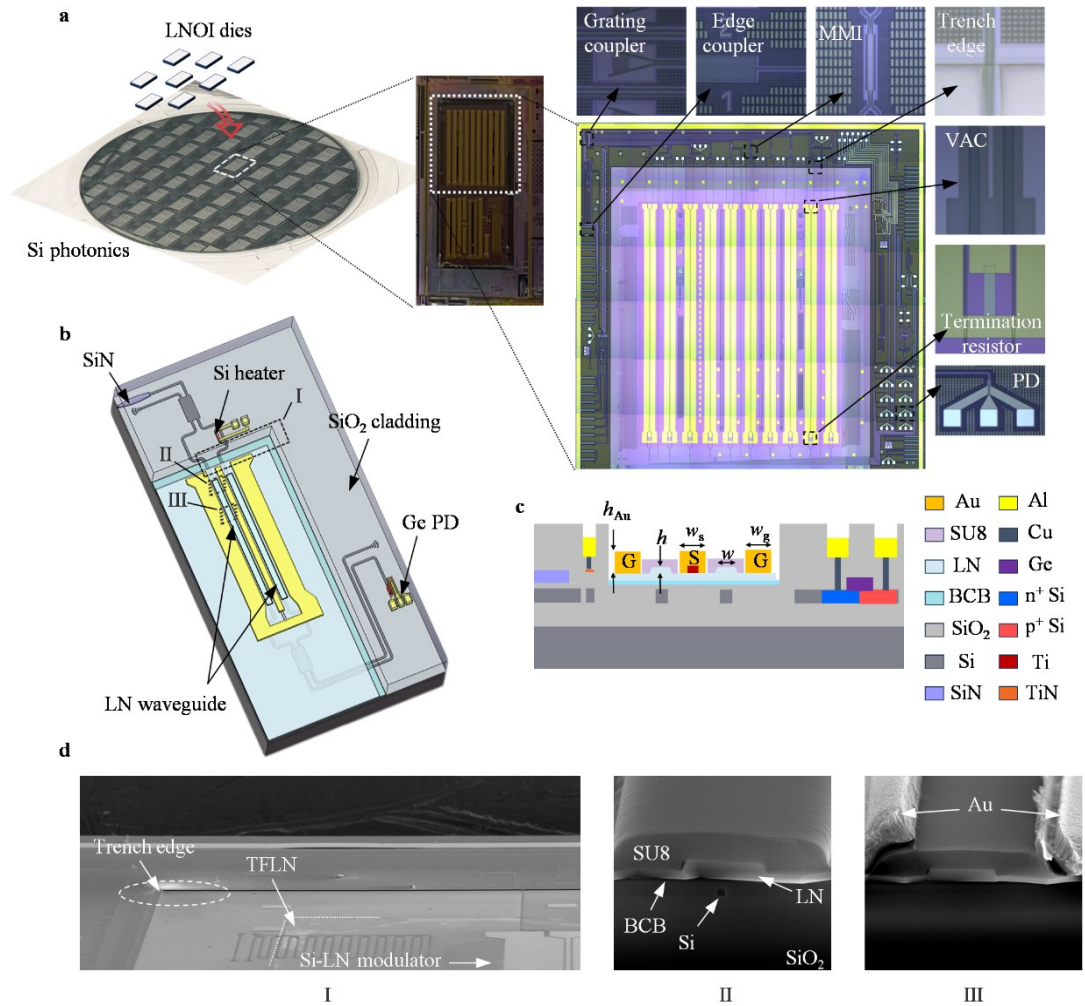

**Fig. 1 Proposed heterogeneous BEOL integration of TFLN on active Si photonics. a,** Schematic of the die-to-wafer bonding process on an 8-inch active Si photonics wafer and optical microscope images of the fabricated chip. The trench region contains the VACs and TFLN modulation sections, while the surrounding Si photonic circuits comprise grating couplers, 3-dB MMI couplers, SiN edge couplers, Si TO phase shifters, and Ge photodetectors. **b,** 3D schematic of a representative transceiver channel. **c,** Cross-sectional view of the heterogeneous integration platform. **d,** Scanning electron microscope images of the trench area (I), the cross-section of VAC (II), and the cross-section of the modulation region (III).

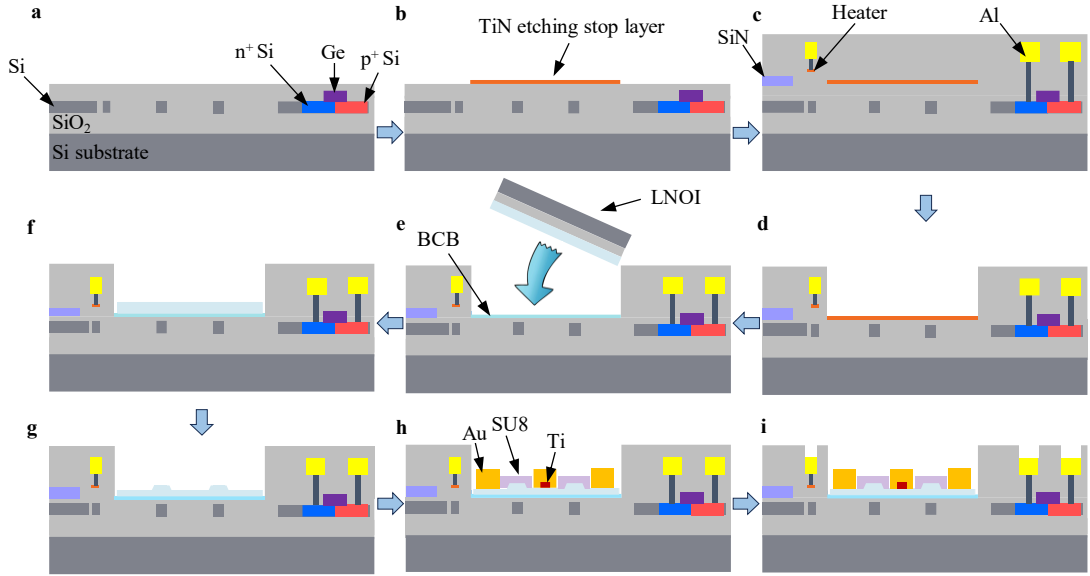

**Fig. 2 Fabrication process of the heterogeneous integration platform. a**, Fabrication of the Si waveguides, Ge epitaxy layer, and Si doping. The wafer was then planarized using $SiO_2$ cladding. **b**, Deposition of the etching stop layer at the modulation region. **c**, Rest of the Si photonics fabrication, including the SiN layer, heaters, and corresponding metallization. **d**, Trench opening for the bonding area. **e**, Removal of the TiN etching stop layer and bonding of LNOI dies. **f**, Removal of the LNOI substrate and oxide layer. **g**, Definition of the TFLN waveguides. **h**, Rest of the TFLN modulation section fabrication, including SU8 over-cladding, Ti termination resistors, and Au electrodes. **i,** Pad opening for heaters and Ge photodetectors.

**Performance of key devices.**
*Interlayer and fiber-to-chip coupling*

Efficient inter-layer and inter-material optical coupling is crucial for photonic heterogeneous integration, which, in the present platform, includes the coupling from the Si waveguides to the TFLN waveguides, the SiN waveguides, and the Ge absorption layer. Evanescent coupling is generally adopted. The VAC connecting the Si and LN layers is carefully engineered to ensure a low loss and a high fabrication tolerance. The Si waveguide is tapered from 450 nm to 180 nm over a length of 200 μm, while the TFLN waveguide remains 1.5 μm wide in the VAC region, and expands to 2.5 μm in the modulation section. The measured coupling efficiency of this VAC exceeds 97%, corresponding to a loss of only ~0.11 dB per coupler (Supplementary Section II). Simulations confirm that tolerances of up to ±300 nm lateral offset and ±20 nm BCB thickness variation can be achieved (Supplementary Section II), which is advantageous for the bonding process of the TFLN dies and the definition of the TFLN patterns in the BEOL technology. A similar strategy applies to Si-SiN coupling, considering the comparable refractive indices of SiN and LN. Here, the Si waveguide is tapered from 450 nm to 160 nm, while the SiN waveguide expands from 350 nm to 1 μm. Owing to the reduced vertical separation (without BCB), as well as a higher CD and overlay accuracy control, the taper length is set to 35 μm. Experimentally, Si-SiN interlayer optical loss as low as 0.06 dB per VAC coupler was achieved (Supplementary Section II). For photodetection, a horizontal PIN structure in Si is adopted with a Ge layer directly grown on Si. The Ge waveguide is also designed in phase matching with the Si waveguide beneath, which ensures a high optical absorption rate over a device length of 55 μm.

In addition, the SiN layer provides an efficient solution for broadband and low-loss coupling to optical fibers. In the present design, the optical mode from the Si waveguide is transferred to the SiN layer via the aforementioned interlayer couplers, and then expanded using an inverse taper from 1 μm to 350 nm width to match the mode field diameter of standard fibers using inverse tapers. Experimentally, a coupling loss of 1.6 dB for TE polarization was achieved from the on-chip Si waveguide to a single-mode fiber with a 7.5 μm mode field diameter over a broad spectral range (Supplementary Section II). Beyond efficient and robust fiber interfaces, the SiN layer also offers a versatile platform for integrating more complex photonic functionalities, such as high-performance wavelength filtering and Kerr nonlinear optics, within this heterogeneous integration platform.

*Modulator*

The modulator adopts a Si-TFLN heterogeneously integrated unbalanced MZI structure, where all passive components and the phase modulation sections were fabricated on Si and TFLN, respectively. Figure 2a shows the measured half-wave voltage ($V_\pi$) under a 100 kHz triangular voltage sweep. The device exhibits a $V_\pi$ of 4.4 V for a 6.4 mm-long modulator, corresponding to voltage–length products ($V_\pi L$) of 2.8 V·cm. Figure 2b shows the normalized wavelength response, indicating a minimal insertion loss of 4 dB and an extinction ratio exceeding 25 dB. The relatively high insertion loss originates from the excess propagation loss in the TFLN waveguides, mainly due to the limited resolution of the contact photolithography adopted here. Using a BEOL stepper, the TFLN waveguide loss can be reduced below 0.27 dB/cm[40]. Considering typical losses of ~0.3×2 dB from the Si MMIs and ~0.11×2 dB from the two VACs, the total insertion loss of the Si-TFLN heterogeneous modulator could be reduced to below 1.5 dB.

Si TO phase shifters were also characterized. Figure 3c shows the transmission versus applied electrical power on the heater. A π-phase shift was achieved with a power consumption ($P_\pi$) of only 17.1 mW on a 1050 Ω heater resistor, owing to the strong TO coefficient of Si. Unlike EO phase shifters on LN, which suffers from a bias drift caused by free-carrier migration, Si based TO phase shifters provide a stable bias control[41]. To confirm this, the modulator was biased at quadrature, and its output was monitored for 30 min without modulation (Fig. 3d). The bias point remained stable, with only minor fluctuations resulted from fiber coupling. These results highlight Si TO phase shifters as a compact and drift-free bias control mechanism in this heterogenous integration platform.

The small-signal electro-optic (EO) response of the modulator was measured using a vector network analyzer (VNA, Keysight N5227B) and a 110 GHz optical component analyzer (Newkey GOCA-110). As shown in Fig. 3e, the 3 dB EO bandwidth of the 6.4 mm-long modulator reaches about 100 GHz. Considering also the half-wave voltage, the performance of this Si-TFLN modulator surpasses that of conventional Si MZI modulators. At low frequencies (<5 GHz), the S21 response shows a slight rise before roll-off, attributed to the intentional mismatch between the ~43 Ω termination resistor and the traveling-wave electrode impedance (see Supplementary Section I). Further improvements are possible with capacitively loaded traveling-wave electrodes and undercut or backside-etched Si substrates[23,42], which can help further enhance the modulation efficiency and bandwidth performances simultaneously.

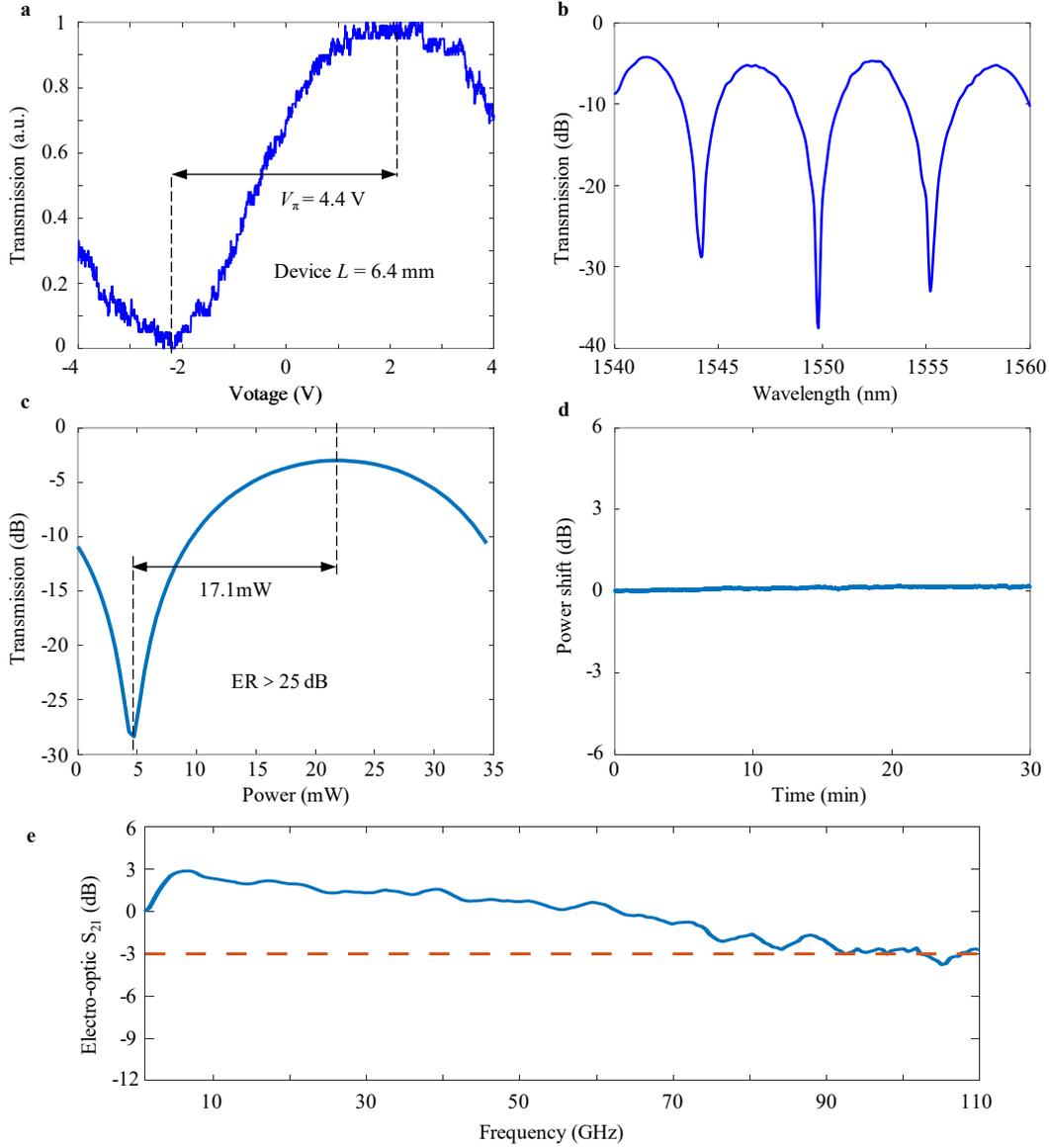

**Fig. 3 Performance of heterogeneous Si-TFLN modulators. a**, Normalized optical transmission as a function of the voltage applied to the traveling-wave electrodes, showing a $V_\pi$ value of 4.4 V. **b**, Normalized optical spectrum showing an insertion loss of 4 dB with an extinction ratio exceeding 25 dB. **c**, Normalized optical transmission as a function of the electrical power applied to the Si TO phase shifter, indicating a $P_\pi$ of 17.1 mW. **d**, Stability of the quadrature bias point set by the TO phase shifter. **e**, EO frequency response ($S_{21}$) of a 6.4-mm-long modulation section, exhibiting a 3-dB bandwidth about 100 GHz.

*Ge photodetector*

The integrated Ge photodetectors were characterized before (pre-bonding) and after (post-bonding) bonding of TFLN. As shown in Figs. 4a and 4b, the current-voltage (IV) characteristics measured at a wavelength of 1550 nm under different optical powers and bias voltages exhibit nearly identical performances, indicating that the bonding process would not introduce any degradation in the Ge photodetectors. The dark current is as low as 55 nA at –1 V, and a responsivity exceeding 0.8 A/W across the whole C band is also measured as shown in Fig. 4c. The small-signal response, measured using an VNA, presents 3-dB bandwidths of 47 GHz and 56 GHz at –1 V and –2 V bias, respectively. These results verify the successful integration of Ge photodetectors within the heterogeneous platform.

*On-chip optical data link*

Enabled by the heterogeneous integration platform, high-performance modulators and photodetectors can be co-integrated on a single Si chip. Using the present technology, we built high-density on-chip optical data links. We further measured the EE bandwidth of the full on-chip data links from the heterogeneous Si-TFLN modulator to the Ge photodetector using an VNA as shown in Fig. 5a, where the Ge photodetectors were biased at –2 V and the wavelength was controlled at the quadrature of the MZI modulators. Figure 5b shows uniform responses across all four channels, with 3 dB bandwidths of ~60 GHz, which is still limited by those of the Ge photodetectors.

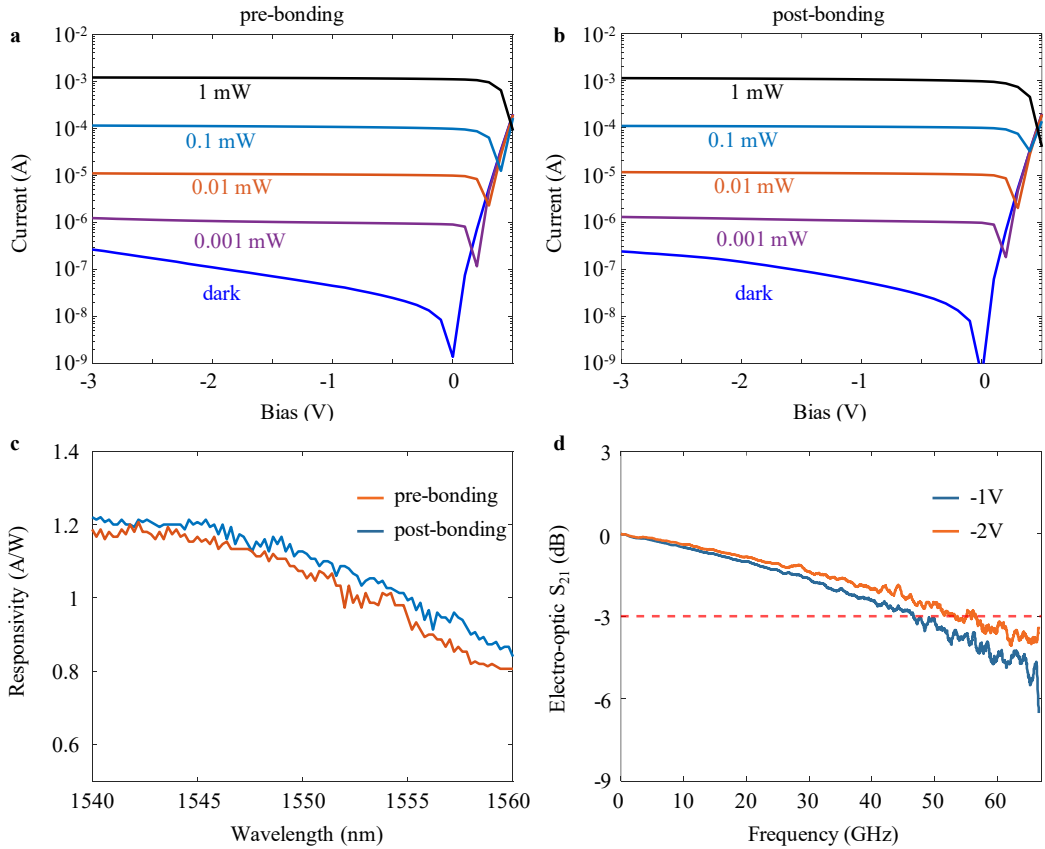

**Fig. 4 Characterization of the Ge photodetector.** IV characteristics of the photodetector (**a**) before and (**b**) after

bonding of TFLN under different optical powers and bias voltages at 1550nm wavelength, showing nearly identical performances. **c**, Responsivity of the photodetector at –1 V bias across the C band. **d**, EO frequency response ($S_{21}$) of the photodetector with 3-dB bandwidths of 47 GHz at –1 V and 56 GHz at –2 V biases.

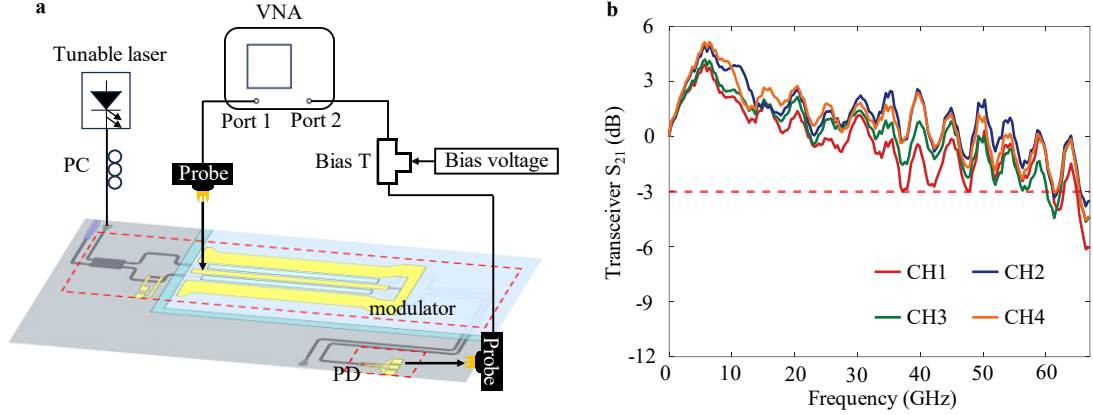

**Fig. 5 EE bandwidth measurement of the heterogeneous integrated on-chip transceiver link. a**, Schematic of the measurement setup. VNA: vector network analyzer, PD: Ge photodetector; PC: polarization controller. **b**, Measured EE $S_{21}$ responses of the four channels under a detector bias of –2 V, exhibiting 3-dB bandwidths of ~60 GHz for all channels.

To verify the feasibility of high-speed optical interconnect on the fabricated chip, on-chip back-to-back data transmission experiments, as illustrated in Fig. 6a, were conducted without any amplifiers in the optical domain. Experimental details are provided in the "Methods" section. Representative results from four channels are shown in Figs. 6b and 6c. For 128 Gbaud OOK signals, bit error rates (BERs) below $2.4 \times 10^{-4}$ were achieved, surpassing the KP4-FEC threshold. For 100 Gbaud PAM4 signals, BERs below $3.8 \times 10^{-3}$ were obtained, satisfying the HD-FEC limit. The BER dependence on received optical power was evaluated by varying the modulator input power with the coupling losses normalized out. Clear eye diagrams shown in Figs. 6b and 6c confirm a high-quality data recovery across all channels. These results demonstrate that high-speed end-to-end data transmission is granted using the fabricated modulators and detectors, and underscore the potential of the present platform for large-capacity optical interconnects and optical NoCs in future wafer-scale computing.

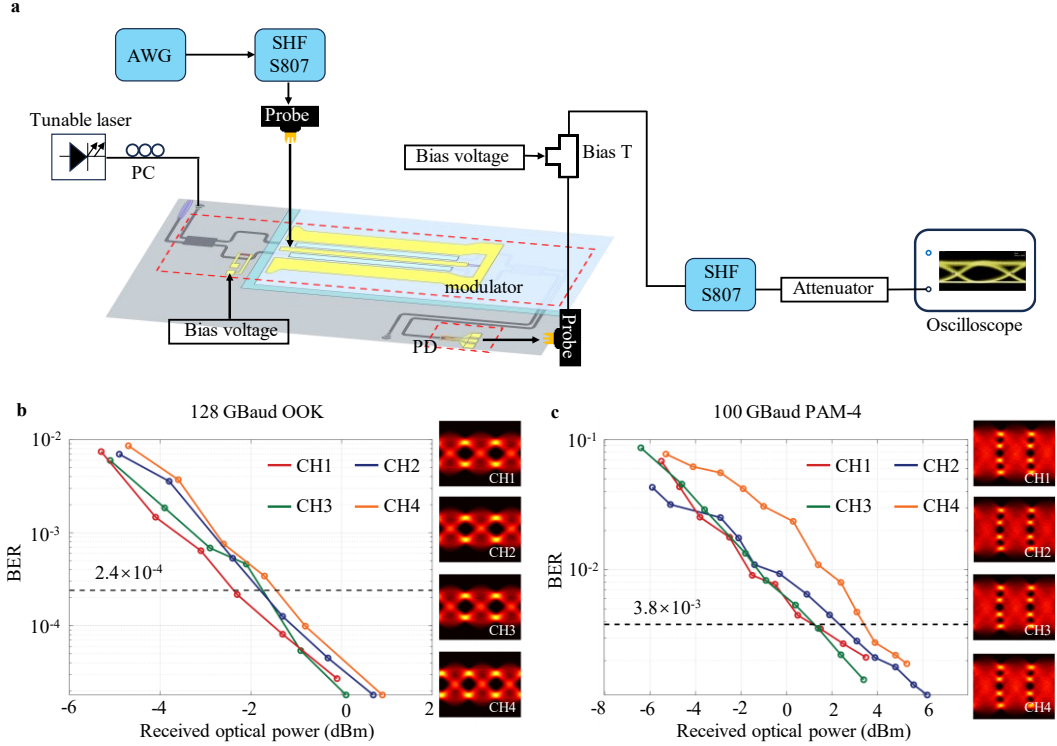

**Fig. 6 Data transmission testing of the on-chip data link. a,** Schematic of the measurement setup. PC: polarization controller, AWG: arbitrary waveform generator, PD: Ge photodetector. Measured curves of BERs versus received optical powers for (**b**) 128 Gbaud OOK and (**c**) 100 Gbaud PAM-4 signals. Eye diagrams are shown at 1 dBm and 4 dBm received optical powers, respectively.

**Discussion and conclusion**

We have demonstrated the first heterogeneous BEOL integration platform that combines TFLN with active silicon photonics, overcoming key challenges in die-to-wafer bonding and multilayer integration. In contrast to previous efforts limited to passive Si circuits, the present platform facilitates to integrate high-speed Ge photodetectors with high-performance TFLN modulators on a single Si chip. Compared with pure Si photonics, TFLN offers intrinsically low loss, superior modulation efficiency, and large bandwidth. The present platform also retains the CMOS compatibility for all the Si photonic components, leverages the mature and highly integrated Si photonics PDKs. This marks a critical step toward scalable, cost-effective, and fully integrated photonic transceiver platforms for data-center optical interconnects and future optical NoCs.

Although the modulators and detectors demonstrated here do not yet reach their ultimate performance limits, there is a clear path towards further improvements. By adopting a capacitively loaded traveling-wave electrode with a removed Si substrate and/or a differential drive, the modulation performance can be enhanced to achieve a modulation figure-of-merit (defined as the 3-dB bandwidth over $V_\pi$ square) up to 100-243GHz/$V^2$ [23,43,44]. With a drive voltage of several volts, the bandwidth of a Si-TFLN modulator could match that of the state-of-the-art Ge photodetectors[6]. Advanced designs such as resonant modulators or slow-light structures could further reduce the TFLN modulator footprint to the submillimeter scale[45-47]. The above facts also indicates that the present heterogeneous integration platform can be extended to even higher data rates.

A further distinctive advantage of this platform lies in its compatibility with SiN, which enables efficient TFLN-Si-SiN coupling within the same chip. This tri-layer integration, combining with high-performance modulation and detection, paves the way for complex system-level functionalities, such as coherent and multi-dimensional multiplexed transceivers, microwave photonics, quantum photonics, and photonic computing.

## Acknowledgements

This work was supported by the National Natural Science Foundation of China (62435016, 62135012), Basic and Applied Basic Research Foundation of Guangdong Province (2024A1515011710), and Guangdong Provincial Key Laboratory of Optical Information Materials and Technology (2023B1212060065).

## Author contributions

## Data availability

The data that support the plots within this paper and other findings of this study are available from the corresponding author upon reasonable request.

## Conflict of interest

The authors declare no competing interests.

## Additional information

**Supplementary information** is available for this paper at xxx

## Methods

### Fabrication

The Si photonics circuits generally adopted the 8-inch 130-nm Si/SiN photonics PDK from CUMEC, China, on a silicon-on-insulator wafer of 220nm thick top Si layer and 3 μm thick buried oxide layer[42]. First, the passive silicon devices, including grating couplers, 3-dB multimode interference (MMI) couplers, and routing waveguides, were fabricated. Subsequently, the Ge photodetectors, as well as the necessary doping regions, were prepared. The wafer surface was then planarized by an $SiO_2$ layer using plasma enhanced chemical vapor deposition and chemical mechanical polishing technologies. The thickness of the $SiO_2$ layer on top of the Si waveguides is ~180 nm. Then, a TiN layer was deposited on the TFLN modulation areas, which served as an etching stop layer for etching the $SiO_2$ trenches afterwards. The SiN tapers for edge couplers were also fabricated on top of this $SiO_2$ layer with a SiN layer thickness of 300 nm. Extra SiO$_2$ claddings, heaters, and other metal structures were subsequently fabricated. In the present wafer, only one metal interconnect layer was used. Nevertheless, more metal layers can be included if necessary. The SiO$_2$ overcladding layer, about 5 μm thick, in the modulation regions was thinned by dry etching until the TiN etching stop layer, which was subsequently removed. At this stage, the wafer fabrication for Si photonics is completed, which in principle can be processed in any ordinary CMOS fabs. Prior to bond the TFLN dies, the surface of the Si photonics wafer was thoroughly cleaned to remove inorganic particles, organic residues, and other contaminants with wet chemicals. An 85-nm-thick BCB adhesive layer was spun onto the wafer. The film was prebaked at 180 °C for 20 min to remove solvents. One or more x-cut LNOI dies (commercially available from NanoLN, China) were then flipped and bonded into the trenches over the modulation regions using a commercial wafer bonder (EVG 501). The bonding was carried out under a force of 600 N in vacuum, followed by an annealing process at 300 °C for 1 h to fully cure the BCB. The silicon substrate and buried oxide of the LNOI dies were subsequently removed by mechanical grinding and selective wet etching, leaving a 500-nm thick TFLN layer attached to the Si photonics wafer. After bonding, the modulation structures, including the TFLN waveguides, traveling-wave electrodes made of Au, and 50 W termination resistors made of Ti, were prepared using ordinary TFLN modulator processes[24]. Here, all the patterning was done using contact i-line photolithography (EVG 620). Finally, the SiO$_2$ cladding above the metal pads for the Ge photodetectors and heaters was etched to expose these pads.

**High-speed data transmission measurement**

The measurement setup for the data transmission is shown in Fig. 6a. A tunable continuous-wave (CW) laser (Agilent 81940A) was used as the optical source. A polarization controller (PC) was used to ensure the transverse-electric polarization before coupling into the chip. Electrical data signals ($2^{19}$ pseudorandom bit sequences) were generated by a 256 GS/s arbitrary waveform generator (AWG) with $V_{pp}$ of 200 mV (Keysight M8199A, 70 GHz analog bandwidth). The signals were further amplified using a broadband microwave amplifier (SHF 807C, 55 GHz bandwidth) before being delivered to the Si-TFLN modulator via a GSG microwave probe (GGB 67A, 67 GHz bandwidth). The modulated optical signals were transmitted through the on-chip waveguides and directly received by the integrated Ge photodetectors. The photodetectors were reverse-biased at −2 V using a bias-tee connected to another microwave probe. The detected microwave signals were amplified by another microwave amplifier, passed through a microwave attenuator, and were recorded by a real-time oscilloscope (Teledyne LeCroy LabMaster 10 Zi-A, 59 GHz bandwidth, 160 GSa/s). Offline digital signal processing, including resampling and feed-forward equalization, was applied to recover the signals and to calculate BER values and eye diagrams.

**References.**